\documentclass[twocolumn,showpacs,floats,floatfix,superscriptaddress,aps,pra]{revtex4}
\usepackage{amsfonts}
\usepackage{amssymb}
\usepackage{amsmath}
\usepackage{calc,graphicx,color,bm}

\def\be{ \begin{equation} }
\def\ee{ \end{equation} }
\def\bea{ \begin{eqnarray} }
\def\eea{ \end{eqnarray} }
\def\bse{ \begin{subequations} }
\def\ese{ \end{subequations} }

\def\i{\,i\,}
\def\e{\,e\,}

\def\to{\rightarrow}

\newcommand{\ket}[1]{\vert #1\rangle}

\def\U{\mathbf{U}}

\def\H{\mathbf{H}}
\def\R{\mathbf{R}}
\def\c{\mathbf{c}}

\def\i{{\emph{i}}}
\def\f{{\emph{f}}}
\def\e{{\emph{e}}}

\def\Re{\,\text{Re}}
\def\Im{\,\text{Im}}

\begin{document}

\author{Boyan T. Torosov}
\affiliation{Institute of Solid State Physics, Bulgarian Academy of Sciences, 72 Tsarigradsko chauss\'{e}e, 1784 Sofia, Bulgaria}
\author{Nikolay V. Vitanov}
\affiliation{Department of Physics, Sofia University, James Bourchier 5 blvd, 1164 Sofia, Bulgaria}
\title{Composite Stimulated Raman Adiabatic Passage}
\date{\today }

\begin{abstract}
We introduce a high-fidelity technique for coherent control of three-state quantum systems, which combines two popular control tools --- stimulated Raman adiabatic passage (STIRAP) and composite pulses.
By using composite sequences of pairs of partly delayed pulses with appropriate phases
 the nonadiabatic transitions, which prevent STIRAP from reaching unit fidelity, can be canceled to an arbitrary order by destructive interference, and therefore the technique can be made arbitrarily accurate.
The composite phases are given by simple analytic formulas, and they are universal for they do not depend on the specific pulse shapes, the pulse delay and the pulse areas.
\end{abstract}


\maketitle

\section{Introduction}
Among the many possibilities for coherent manipulation of quantum systems, stimulated Raman adiabatic passage (STIRAP) is one of the most widely used and studied \cite{STIRAP}.
This technique transfers population adiabatically between two states $\ket{1}$ and $\ket{3}$ in a three-state quantum system, without populating the intermediate state $\ket{2}$ even when the time-delayed driving fields are on exact resonance with the respective pump and Stokes transitions.
The technique of STIRAP relies on the existence of a dark state, which is a time-dependent coherent superposition of the initial and target states only, and which is an eigenstate of the Hamiltonian if states $\ket{1}$ and $\ket{3}$ are on two-photon resonance.
Because STIRAP is an adiabatic technique, it is robust to variations in most of the experimental parameters.

In the early applications of STIRAP in atomic and molecular physics its efficiency, most often in the range 90-95\%, has barely been scrutinized because such an accuracy suffices for most purposes.
Because STIRAP is resilient to decoherence linked to the intermediate state (which is often an excited state) this technique has quickly attracted attention as a promising control tool for quantum information processing \cite{STIRAP-QIP}.
The latter, however, demands very high fidelity of operations, with the admissible error at most $10^{-4}$, which is hard to achieve with the standard STIRAP because, due to its adiabatic nature, it approaches unit efficiency only asymptotically, as the temporal pulse areas increase.
For usual pulse shapes, e.g., Gaussian, the necessary area for the $10^{-4}$ benchmark is so large that it may break various restrictions in a real experiment.

Several scenarios have been proposed to optimize STIRAP in order to achieve such an accuracy.
Because the loss of efficiency in STIRAP derives from incomplete adiabaticity, Unanyan \emph{et al.} \cite{Unanyan97}, and later Chen \emph{et al.} \cite{Chen10}, have proposed to annul the nonadiabatic coupling by adding a third pulsed field on the transition $\ket{1}\to\ket{3}$.
However, this field must coincide in time with the nonadiabatic coupling exactly; its pulse area, in particular, must equal $\pi$, which makes the pump and Stokes fields largely redundant.
An alternative approach to improve adiabaticity is based on the Dykhne-Davis-Pechukas formula \cite{DDP},
 which dictates that nonadiabatic losses are minimized when the eigenenergies of the Hamiltonian are parallel.
This approach, however, prescribes a strict time dependences for the pump and Stokes pulse shapes \cite{Chen12},
 or for both the pulse shapes and the detunings \cite{Dridi}.

Another basic approach to robust population transfer, which is an alternative to adiabatic techniques, is the technique of composite pulses, which is widely used in nuclear magnetic resonance (NMR) \cite{NMR}, and more recently, in quantum optics \cite{QO,CAP}.
This technique, implemented mainly in two-state systems, replaces the single pulse used traditionally for driving a two-state transition by a sequence of pulses with appropriately chosen phases; these phases are used as a control tool for shaping the excitation profile in a desired manner, e.g., to make it more robust to variations in the experimental parameters --- intensities and frequencies.
Recently, we have proposed a hybrid technique --- composite adiabatic passage (CAP) --- which combines the techniques of composite pulses and adiabatic passage via a level crossing in a two-state system \cite{CAP}.
CAP can deliver extremely high fidelity of population transfer, far beyond the quantum computing benchmark, and far beyond what can be achieved with a single frequency-chirped pulse.
Recently, the CAP technique has been demonstrated experimentally in a doped solid \cite{Schraft}.

\begin{figure}[tbph]
\centering \includegraphics[width=0.95\columnwidth]{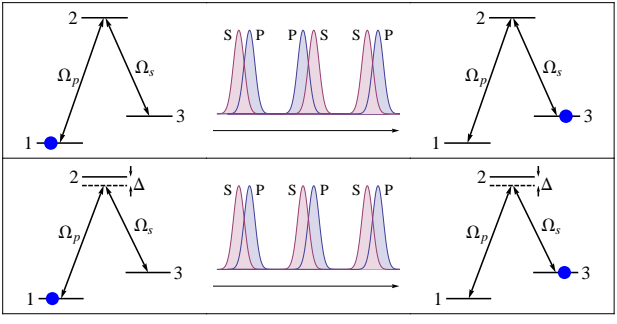}
\caption{Composite STIRAP.
The population is transferred from state $\ket{1}$ to state $\ket{3}$ via a sequence of pump-Stokes pulse pairs.
On one-photon resonance (top), the order of the pump and Stokes pulses is reversed from pair to pair,
 while off single-photon resonance it is the same for all pulse pairs.
}\label{tog}
\end{figure}

In this paper, we combine the two basic techniques --- of composite pulses and STIRAP --- into a hybrid technique, which we name \emph{composite STIRAP}.
This technique, which represents a sequence of an odd number of forward and backward ordinary STIRAPs,
$\ket{1}\to\ket{3}\to\ket{1}\to\ket{3}\to\cdots\to\ket{1}\to\ket{3}$, adds to STIRAP the very high fidelity of composite pulses.
Each individual STIRAP can be very inaccurate, the affordable error being as much as 20-30\%, but all errors interfere destructively and cancel in the end, thereby producing population transfer with a cumulative error far below the quantum computing benchmark of $10^{-4}$.
We derive an analytical formula for the composite phases, applicable to an arbitrary odd number of pulse pairs $N$; the phases do not depend on the shape of the pulses and their mutual delay.

The dynamics of a three-state $\Lambda$ system (Fig. \ref{tog}) is described by the Schr\"{o}dinger equation,
\be\label{Schr}
\i \hbar\partial_t \mathbf{c}(t) = \H(t)\mathbf{c}(t),
\ee
where the vector $\c(t) = [c_1(t), c_2(t), c_3(t)]^T$ contains the three probability amplitudes.
The Hamiltonian in the rotating-wave approximation and on two-photon resonance between states $\ket{1}$ and $\ket{3}$ is
\be\label{H}
\H(t) = \frac\hbar 2 \left[ \begin{array}{ccc} 0  & \Omega_{p}(t) & 0 \\
\Omega^{\ast}_{p}(t)  & 2\Delta - \i\gamma & \Omega_{s}(t)\\
0 & \Omega^{\ast}_{s}(t)& 0
\end{array} \right],
\ee
where $\Omega_{p}(t)$ and $\Omega_{s}(t)$ are the Rabi frequencies of the pump and Stokes fields, $\Delta$ is the one-photon detuning between each laser carrier frequency and the Bohr frequency of the corresponding transition, and $\gamma$ is the population loss rate from state $\ket{2}$; we assume $\gamma=0$.
States $\ket{1}$ and $\ket{3}$ are coupled by $\Omega_{p}(t)$, while states $\ket{2}$ and $\ket{3}$ are coupled by $\Omega_{s}(t)$.
The evolution of the system is described by the propagator $\U$, which connects the amplitudes at the initial and final times, $t_\i$ and $t_\f$: $\c(t_\f)=\U(t_\f,t_\i)\c(t_\i)$.
The mathematics is substantially different when the pump and Stokes fields are on resonance or far off-resonance with the corresponding transition:
therefore we consider these cases separately.

\section{Resonant STIRAP}
First, we will consider the one-photon resonance, $\Delta=0$.
Then there is a mapping between the three-state problem and a corresponding two-state problem \cite{Feynman, NVV&BWS} described by the Hamiltonian
\be
\H(t) = \frac\hbar 2 \left[ \begin{array}{cc} -\Omega_{s}(t)  & \Omega_{p}(t) \\
\Omega_{p}(t)  & \Omega_{s}(t)
\end{array} \right].
\ee
(In this correspondence, $\Omega_{p}(t)$ and $\Omega_{s}(t)$ are assumed real.)
In general, if the two-state propagator is parameterized in terms of the complex Cayley-Klein parameters $a$ and $b$ ($|a|^2+|b|^2=1$) as
\be
\U = \left[\begin{array}{cc}  a  & b \\  -b^{\ast} & a^{\ast} \end{array} \right],
\ee
 we can write the propagator of STIRAP as
\be\label{U-ab}
\U =  \left[ \begin{array}{ccc} \left| a \right|^2-\left| b \right|^2  & -2 \i \Im \left(a b^{\ast}\right)  & 2 \Re \left(a b^{\ast}\right) \\
2 \i \Im \left(a b\right)  & \Re \left( a^2 +b^2 \right) & -\i \Im \left( a^2 -b^2 \right)\\
-2 \Re \left(a b\right) & -\i \Im \left( a^2 +b^2 \right) & \Re \left( a^2 -b^2 \right)
\end{array} \right].
\ee
If $\Omega_{p}(t)$ and $\Omega_{s}(t)$ are reflections of each other,
$\Omega_{p}(t)=\Omega_{s}(\tau-t)$ [e.g., if $\Omega_p(t)$ and $\Omega_s(t)$ are identical symmetric functions of time], where $\tau$ is the pulse delay, then it is easily shown that $\Im a = -\Im b$.
We use this property to parameterize the STIRAP propagator \eqref{U-ab} as
\bse
\begin{align}
&a=\cos\theta\cos\phi + \frac{\i}{\sqrt{2}}\sin\theta, \\
&b=\cos\theta\sin\phi - \frac{\i}{\sqrt{2}}\sin\theta.
\end{align}
\ese
In the adiabatic limit, $\Re \left(a b\right)=1/2$;
 hence $\theta=\pi/2$.

\begin{table}[tb]
\caption{Pump and Stokes phases for different number of pulse pairs $N$ for resonant composite STIRAP.}
\label{table1}
\begin{tabular}{ll}
\hline
\hline
$N$& Phases $(\alpha_1,\beta_1;\alpha_2,\beta_2; \dots; \alpha_N,\beta_N)$ \\
\hline
$3$&   $(0, 1; 3, 3; 1, 0)\pi/3$\\
$5$&   $(0, 4; 5, 8; 3, 3; 8, 5; 4, 0)\pi/5$ \\
$7$&   $(0, 9; 7, 1; 5, 8; 12, 12; 8, 5; 1, 7; 9, 0)\pi/7$ \\
$9$&   $(0, 16;9, 6;7, 15;16, 3;12, 12;3, 16;15, 7;6, 9;16 , 0)\pi/9$ \\
\hline
\hline
\end{tabular}
\end{table}

For backward STIRAP from state $\ket{3}$ to state $\ket{1}$, we need to exchange the order of the pump and Stokes pulses.
The corresponding propagator is
\be
\widetilde{\U}=\R\U\R,\qquad \R=\left[ \begin{array}{ccc} 0  & 0 & 1 \\
0  & 1 & 0\\
1 & 0 & 0
\end{array} \right],
\ee
A constant phase shift in the Rabi frequencies, $\Omega_{p}(t)\to\Omega_{p}(t)\e^{\i\alpha}$ and $\Omega_{s}(t)\to\Omega_{s}(t)\e^{\i\beta}$, is
imprinted into the propagator as
\be
\U_{\alpha,\beta} = \Phi\U\Phi^{\ast},\qquad \Phi =  \left[ \begin{array}{ccc} \e^{\i\alpha}  & 0 & 0 \\
0 & 1 & 0\\
0 & 0 & \e^{-\i\beta}
\end{array} \right].
\ee
A sequence of $N$ STIRAPs (where $N$ is an odd number), each with phases $\alpha_k$ and $\beta_k$, produces the propagator
\be\label{U^N}
\U^{(N)} = \U_{\alpha_N,\beta_N} \widetilde{\U}_{\alpha_{N-1},\beta_{N-1}} \cdots \U_{\alpha_3,\beta_3} \widetilde{\U}_{\alpha_2,\beta_2}\U_{\alpha_1,\beta_1}.
\ee
Next we expand the propagator elements $\U^{(N)}_{11}$ and $\U^{(N)}_{21}$ around $\theta=\pi/2$ and find the phases which nullify as many terms in the expansions as possible.
We have thereby derived the following analytic formula for the composite-STIRAP phases:
\bse\label{phases}
\begin{align}
&\alpha_k^{(N)}= \pi \left\lfloor \frac{k}{2} \right\rfloor - \frac{\pi}{N} \left\lfloor \frac{k-1}{2}  \right\rfloor\left( 1 + \left\lfloor
\frac{k-1}{2} \right\rfloor \right), \\
&\beta_k^{(N)}=\alpha_{N+1-k}^{(N)},
\end{align}
\ese
where $k=1,2\dots,N$. The first few cases are explicitly shown in Table \ref{table1}.
Since no assumptions are made about the Cayley-Klein parameters in the derivation, the composite phases \eqref{phases} do not
depend on the pulse shapes, the pulse delay and the pulse areas.
We note here that, since these phases are solutions of a system of nonlinear algebraic equations, other solutions also exist;
 they, however, produce the same results as the set \eqref{phases}.
Moreover, a common shift in the pump (or/and Stokes) phases does not change the fidelity but causes only a phase shift in the probability amplitudes.

\begin{figure}[t]
\centering \includegraphics[width=1.0\columnwidth]{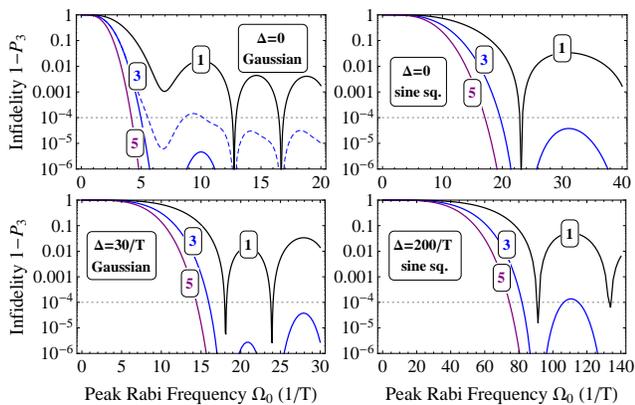}
\caption{Infidelity $1-P_3$ as a function of the peak Rabi frequency for a single STIRAP, compared with three and five-component composite STIRAP.
The pulse shapes are given by Eqs. \eqref{gaussians} (left frames) and \eqref{sinsq} (right frames).
The composite phases are given by Eq.~\eqref{phases} for resonant STIRAP (upper frames) and Eq.~\eqref{CAP-phases} for far-off-resonant STIRAP (lower frames). The dashed curve in the upper left frame is the fidelity of composite STIRAP when a random error of 1\% is included in the phases.
}
\label{c-STIRAP}
\end{figure}

\begin{figure}[t]
\centering \includegraphics[width=0.9\columnwidth]{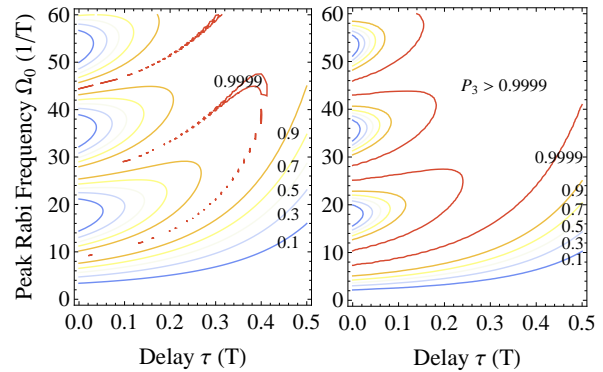}
\caption{Final population $P_3$ as a function of the pulse delay and the peak Rabi frequency for a single resonant STIRAP (top) and a sequence
of five resonant STIRAPs (bottom), with phases given by Eq.~\eqref{phases}, for $\sin^2$ pulse shapes,  Eqs.~\eqref{sinsq}.}
\label{ContRes}
\end{figure}

In Fig. \ref{c-STIRAP} we compare the efficiency of single STIRAP with composite STIRAP for $N=3$ and 5.
We assume that the pump and Stokes pulses share the same shape, which we take to be either Gaussian,
\be\label{gaussians}
\Omega_p=\Omega_0 e^{-(t-\tau/2)^2/T^2},\quad
\Omega_s=\Omega_0 e^{-(t+\tau/2)^2/T^2},
\ee
or sine squared,
\bse\label{sinsq}
\begin{align}
&\Omega_p=\Omega_0\sin^2\left(\pi\frac{t-\tau}{T}\right),\quad t \in \left[\tau , T+\tau\right],\\
&\Omega_s=\Omega_0\sin^2\left(\pi \frac{t}{T}\right),\quad t \in \left[0 , T \right]
\end{align}
\ese
where $T$ is the pulse width and $\tau$ is the delay between the pulses.
We take a delay $\tau=T$ for Gaussian shapes and $\tau=T/\pi$ for sin$^2$ shapes in the simulations.
We see in Fig. \ref{c-STIRAP} that even a sequence of three STIRAPs is enough to achieve extremely high fidelity with an error below $10^{-6}$, which is impossible with a single STIRAP, unless we use huge pulse areas, far outside the axis range.
The robustness of the method is seen in Fig.~\ref{ContRes}, which compares the fidelity of single STIRAP and composite STIRAP with $N=5$.
The high-fidelity region with error below $10^{-4}$ of composite STIRAP is hugely expanded compared to single STIRAP.

\section{Nonresonant STIRAP}
We now focus on the nonresonant case of Hamiltonian \eqref{H}, $\Delta\neq 0$.
If the detuning is small, $\Delta T\ll 1$, than the composite phases do not deviate much from the resonant formulae \eqref{phases} and we can still use them.
However, if the detuning gets larger, then the composite phases depend on $\Delta$;
 an exact formula for the phases does not appear to exist and their values are to be calculated numerically.
When the detuning is very large, we can eliminate adiabatically state $\ket{2}$ and we are left with an effective (symmetric) coupling
$\Omega_{\text{eff}}=-\Omega_p\Omega_s/2\Delta$ between states $\ket{1}$ and $\ket{3}$ and an effective (antisymmetric) detuning
$\Delta_{\text{eff}}=(|\Omega_p|^2-|\Omega_s|^2)/ 2\Delta$.
This effective two-state problem reduces to the already studied CAP technique \cite{CAP}, where the phases are known and an
analytical formula also exists,
\be\label{CAP-phases}
\alpha_k^{(N)} - \beta_k^{(N)} = \left( N+1-2\left\lfloor \frac{k+1}{2} \right\rfloor \right)\left\lfloor \frac{k}{2} \right\rfloor \frac \pi N.
\ee
The values for the first few cases are given in Table \ref{table2}.

\begin{table}[tb]
\caption{Phases for different number $N$ of pulse pairs for off-resonant composite STIRAP.
Because only the phase difference $\alpha_k-\beta_k$ between the pump and Stokes phases is important, we set all Stokes phases $\beta_k=0$ and show the pump phases $\alpha_k$ only ($k=1,2,\ldots,N$).}
\label{table2}
\begin{tabular}{ll}
\hline
\hline
$N$& Phases $(\alpha_1,\alpha_2, \dots \alpha_N)$ \\
\hline
$3$&   $(0, 1, 0)2\pi/3$\\
$5$&   $(0, 2,1,2, 0)2\pi/5$ \\
$7$&   $(0, 3,2,4,2,3, 0)2\pi/7$ \\
$9$&   $(0, 4,3,6,4,6,3,4 , 0)2\pi/9$ \\
\hline
\hline
\end{tabular}
\end{table}

\begin{figure}[t]
\centering \includegraphics[width=0.9\columnwidth]{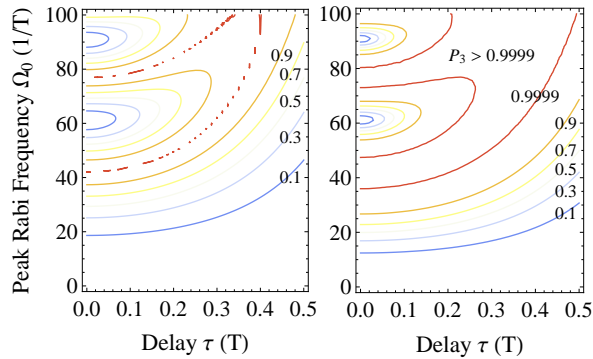}
\caption{Final population $P_3$ as a function of the pulse delay and the peak Rabi frequency for a single off-resonant STIRAP (top) and a sequence of five STIRAPs (bottom), with phases given by Eq.~\eqref{CAP-phases}, for $\sin^2$ pulse shapes,  Eqs.~\eqref{sinsq}. The detuning is $\Delta=100/T$.}
\label{ContOffRes}
\end{figure}

The fidelity of the nonresonant composite STIRAP is illustrated in Fig.~\ref{c-STIRAP} (bottom frames) and Fig.~\ref{ContOffRes}.
Again, composite STIRAP greatly outperforms single STIRAP in terms of fidelity and robustness.

\section{Discussion}
Composite STIRAP may be affected by several sources of errors.
In the first place, errors in the composite phases should be held low in order to keep the high fidelity.
We found that an error below 1\% in the phases, which is relatively easy to achieve in the lab, can be tolerated.
In Fig. \ref{c-STIRAP} we have added a curve, which demonstrates the fidelity of composite STIRAP for $N=3$ and a standard deviation of 0.01 radians in the composite phases \footnote{The distribution of the phases is assumed normal with a standard deviation of 0.01 radians and the curve is calculated after averaging a large number of Monte Carlo simulations.};
 despite this error, the technique still has ultrahigh fidelity, with an error below $10^{-4}$.

\begin{figure}[t]
\centering \includegraphics[width=0.85\columnwidth]{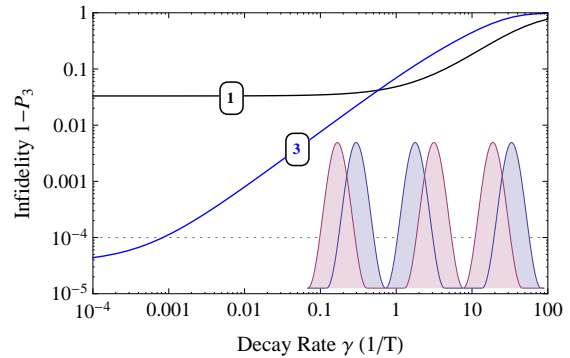}
\caption{Infidelity versus decay rate. We assume $\sin^2$ pulse shapes with $\Omega_0 =30/T$ and $\Delta=0$.}
\label{decay}
\end{figure}

STIRAP owes much of its great popularity to the fact that it can operate, unlike other techniques, in the presence of population losses from the middle state $\ket{2}$.
However, when ultrahigh fidelity is aimed the presence of such losses can reduce the fidelity and they cannot be very large.
(The decay can be harmful only in the resonant case, while off-resonant composite STIRAP is much more resilient to them.)
We have found that in the resonant case, if the decay rate is sufficiently low, or if the pulse duration is sufficiently short ($\gamma T \lesssim 1$), composite STIRAP still maintains high fidelity and outperforms the standard STIRAP, as seen in Fig.~\ref{decay}.
As $\gamma$ increases above $1/T$, STIRAP behaves better but the fidelities of both STIRAP and composite STIRAP drop rapidly and are inadequate for quantum computing purposes.
The presence of losses can be compensated with higher Rabi frequency; as a rough estimate the scaling law $\Omega_0\propto\sqrt{\gamma}$ applies.
It is also important to note that in the presence of decay the pulse pairs should be as close to each other as possible, as in the inset of Fig.~\ref{decay}.
This is readily achieved with microsecond and nanosecond pulses, e.g., produced by acoustooptic modulators,
  as has been demonstrated recently in a doped-solid experiment \cite{Beil11}.

Because composite STIRAP involves $N$ pulse pairs, its duration is longer than STIRAP by the same factor, given that there are no gaps between the pulse pairs, as shown in the inset of Fig.~\ref{decay}.
In return, composite STIRAP gives a fidelity which cannot be achieved with ordinary STIRAP, even with the much higher pulse areas.
Thus the main advantage of composite STIRAP over ordinary STIRAP is the ultrahigh fidelity.
The main advantage of composite STIRAP over other variations of STIRAP, which provide ``shortcuts'' to adiabaticity by eliminating or reducing the nonadiabatic coupling \cite{Unanyan97,Vasilev,Dridi,Chen10,Chen12} is the simplicity of implementation, which requires just the control of the relative phases between the pulse pairs, and the preserved robustness of STIRAP with respect to variations in the interaction parameters.
The ``shortcuts'' techniques use less pulse area, and therefore are faster than composite STIRAP (although still slower than resonant techniques which use areas of just $\pi\sqrt{2}$ \cite{Vitanov98}), but they give away most of the robustness of STIRAP by imposing strict restrictions on the pulse shapes, and some of them on the detunings too; some of them even place considerable transient population in the intermediate state.

\section{Conclusion}
The hybrid technique proposed here combines two popular methods for manipulation of quantum systems --- STIRAP and composite pulses.
It greatly outperforms the standard STIRAP in terms of fidelity due to cancelation of the nonadiabatic errors by destructive interference.
The greatly enhanced fidelity, well beyond the quantum computing benchmark, while preserving STIRAP's robustness against variations in the interaction parameters, makes composite STIRAP a promising technique for quantum information processing.


\section*{Acknowledgments}
This work is supported by the Bulgarian NSF grants D002-90/08 and DMU-03/103, and the Alexander von Humboldt Foundation.


\end{document}